\documentstyle[12pt]{article}
\begin{document}
\hfill
NIKHEF-H/95-027

\hfill
June 1995

\begin{center}
{\huge \bf The large top quark mass expansion for Higgs boson
decays into bottom quarks and into gluons  }\\ [8mm] S.A.
Larin$^{ab}$, T. van Ritbergen$^b$, J.A.M. Vermaseren$^b$ \\ [3mm]
\begin{itemize}
\item[$^a$]
 Institute for Nuclear Research of the
 Russian Academy of Sciences   \\
 60th October Anniversary Prospect 7a,
 Moscow 117312, Russia
\item[$^b$]
NIKHEF-H, P.O. Box 41882, \\ 1009 DB, Amsterdam, The Netherlands \\
\end{itemize}
\end{center}
\begin{abstract}
We calculate the
large top quark mass expansions for the
$H\rightarrow b \overline{b}$ decay rate in the order $\alpha_{s}^2$
and for the $H\rightarrow$ gluons decay rate in the order
$\alpha_{s}^3$.  The obtained expansions rapidly converge in the
region of their validity, $M_H < 2 m_t$, i.e. below the threshold of
$t \overline{t} $ production.
\end{abstract}

\section{Introduction}
With the discovery of the top quark
\cite{cdfd0}, the Higgs boson remains as the only fundamental particle
of the Standard Model which is not found experimentally. Considerable
efforts were devoted to calculate perturbative corrections within
the Standard Model to different Higgs boson decay channels, for a review see
\cite{kniehlrep}. In the present paper we calculate top quark
mass corrections to Higgs boson decays into bottom quark-antiquark
pairs and into gluons, i.e. to the
Higgs boson partial widths $\Gamma(H\rightarrow b\bar{b})$
and $\Gamma(H\rightarrow $gluons).

The decay process
$H\rightarrow b\bar{b}$ is the dominant decay channel for the Higgs
boson with an intermediate mass $M_H<2M_W$ and will be of prime
importance in the future experimental searches of the Higgs boson
at colliders.

The decay channel $H\rightarrow$ gluons is
interesting since heavy quarks that mediate this process
contribute to the decay rate
without being suppressed by their large mass which
eventually may provide a way
to count the number of heavy quarks beyond the Standard Model.

The purpose of the present paper is to
examine the convergence properties of the
series of top quark mass
suppressed corrections to the Higgs boson decay rate in the region
$M_H<2m_t$, where the large top quark mass expansion can be applied.
 These corrections a priori can be sizable but they turn
out to be small for the processes considered so far.  In ref.
\cite{zbbig} we considered the top quark mass corrections to the Z
boson and $\tau$ lepton decays into hadrons.  In this letter we
continue this program to the case of the Higgs decays into hadrons.

Throughout
this paper in our calculations we use
dimensional regularization \cite{dimreg}
and the standard modification
of the minimal subtraction scheme \cite{ms}, the $\overline{MS}$
scheme \cite{msbar}.

\section{The order $\alpha_{s}^2 $ corrections to
$\Gamma(H\rightarrow b \overline{b})$ }

We use the optical theorem to calculate
the top mass suppressed corrections to the
partial decay width $\Gamma(H\rightarrow b \overline{b})$
in the order $\alpha_{s}^2$. To this end we should
calculate the imaginary parts of the 3-loop Higgs boson propagator
diagrams of non-singlet and singlet types
with top quark loops listed
in fig.1a and fig.1b correspondingly.

%\[ \epsfxsize=11.5cm \epsfbox{higgs1.eps} \]
\[ \vspace{3cm} \]
\[ \parbox{14cm}{\small
{\bf Figure 1.} Diagrams contributing to top mass suppressed corrections
 to the partial decay width $\Gamma(H\rightarrow b \overline{b})$.
 Thin lines indicate bottom quark propagators, thick lines indicate
 top quark propagators and spiral lines indicate gluons. The symbol
 $\otimes$ indicates a Yukawa type Higgs--quark vertex. Diagrams
 1a are of the non-singlet type, Diagram 1b is of the singlet type.}
\vspace{.2cm} \]
To extract the contribution to
this channel from the singlet "double triangle" diagram in fig.1b we
should subtract from the imaginary part of this diagram the
contribution to $\Gamma(H\rightarrow$ gluons) associated with two
gluon cut in this singlet diagram.  The contribution from this two
gluon cut can be straightforwardly calculated \cite{twogluoncut} by
the use of Feynman parameters.

For the calculation of the propagator diagrams in fig.1 we
applied the diagrammatic large top quark mass expansion to each of the
diagrams separately, using the technique which was previously
used in \cite{zbbig} and which is based on the method developed in
\cite{gorishny}.
Although the formal parameter of this expansion is not small (the
mass ratio $M_H/m_t$ can be larger than 1) this
expansion is applicable below the threshold for
the production of $t \overline{t}$ pairs, i.e. $M_H < 2 m_t$.
(One can say that the real expansion parameter is $M_H / 2 m_t$.)
For the actual calculations we relied on the symbolic manipulation
program FORM\cite{form} and we used the package MINCER\cite{mincer}
together with additional FORM routines to perform massive integrals.

The Yukawa type Higgs--quark couplings contribute a factor $m_b^2$
for the non-singlet diagrams and a factor $m_b m_t$ for the singlet
diagram.  We work in the leading order in small b-quark mass, $m_b <<
M_H$, and nullify other light quark masses.  This technically means
that for the non-singlet diagrams the mass in the b-quark  propagators
is nullified from the start but for the singlet diagram one power of
the b-quark mass from the propagators must be kept to prevent the
trace over the Dirac matrices in the b-triangle from vanishing.
The higher small $m_b^2$-corrections can be found in
\cite{surg} for non-singlet contributions and in
\cite{bbbarsin2} for the full case.

After addition of the  massless ($m_t$ independent)
non-singlet contributions calculated in \cite{bbmassless}
we obtain the following result for the large top quark mass expansion
of the Higgs decay rate into $b\overline{b}$ in the leading order of the
b-quark mass
\[ \Gamma(H\rightarrow b\overline{b})=\Gamma^{NS}(H\rightarrow b\overline{b})
      +\Gamma^{S}(H\rightarrow b\overline{b}) , \hspace{9cm}\]
 \[
\Gamma^{NS}(H\rightarrow b\overline{b}) =
\frac{G_{F}M_{H}(m_b^{(6)})^2}{4\pi\sqrt{2}}
  n
\left[ 1+ \left( \frac{\alpha_{s}^{(6)}}{\pi} \right)  f_{1}^{NS,0} \right. +
 \hspace{7cm} \]
\begin{equation} \label{bbnsfull}
 \left. \hspace{4cm}
\left( \frac{\alpha_{s}^{(6)}}{\pi} \right)^{2}
\left( f_{2}^{NS,0} + f_{2}^{NS,1}\frac{M_H^2}{m_{t}^2}+
             f_{2}^{NS,2}\frac{M_H^4}{m_{t}^4}
          + f_{2}^{NS,3}\frac{M_H^6}{m_{t}^6}
     +O(\frac{M_H^8}{m_{t}^8}) \right) \right] ,
\end{equation}
$ f_{1}^{NS,0} = C_{F}  ( \frac{17}{4} ) $,
\\ \\
$ f_{2}^{NS,0} = C_F^2 \left[
         \frac{691}{64} - \frac{9}{4} \zeta_3 - \frac{3}{8} \pi^2
       - \frac{105}{16} \ln(\frac{M_H^2}{\mu^2})
       + \frac{9}{8} \ln^2(\frac{M_H^2}{\mu^2}) \right] $

$ +  C_A C_F  \left[
        \frac{893}{64} - \frac{31}{8} \zeta_3 - \frac{11}{48} \pi^2
         - \frac{71}{12} \ln(\frac{M_H^2}{\mu^2})
         + \frac{11}{16} \ln^2(\frac{M_H^2}{\mu^2}) \right]$

$+ T_F C_F N_f \left[
         - \frac{65}{16} + \zeta_3 + \frac{1}{12} \pi^2
             + \frac{11}{6}\ln(\frac{M_H^2}{\mu^2})
           - \frac{1}{4} \ln^2(\frac{M_H^2}{\mu^2}) \right] $

$+ T_F C_F \left[
         \frac{337}{72} - \zeta_3  - \frac{1}{12} \pi^2
       -\frac{11}{6} \ln(\frac{M_H}{m_t^2})
       + \frac{1}{4} \ln^2(\frac{M_H^2}{m_t^2})  \right] $,
\\ \\
$ f_{2}^{NS,1} = T_F C_F \left[ \frac{107}{450}
           - \frac{1}{15} \ln(\frac{M_H^2}{m_t^2}) \right]$,
\\ \\
$ f_{2}^{NS,2} = T_F C_F \left[ - \frac{529}{58800}
            + \frac{1}{280} \ln(\frac{M_H^2}{m_t^2}) \right]$,
\\ \\
$ f_{2}^{NS,3} = T_F C_F \left[ \frac{2719}{3572100}
            - \frac{1}{2835} \ln(\frac{M_H^2}{m_t^2}) \right]$,
\\ \\
\begin{equation} \label{bbsfull}
\Gamma^{S}(H\rightarrow b\overline{b}) =
\frac{G_{F}M_{H}(m_b^{(6)})^2}{4\pi\sqrt{2}}
  n
\left[
\left( \frac{\alpha_{s}^{(6)}}{\pi} \right)^{2}
\left( f_{2}^{S,0} + f_{2}^{S,1}\frac{M_H^2}{m_{t}^2}+
             f_{2}^{S,2}\frac{M_H^4}{m_{t}^4}
      +  f_{2}^{S,3}\frac{M_H^6}{m_{t}^6}
     +O(\frac{M_H^8}{m_{t}^8}) \right) \right] ,
\hspace{4cm} \end{equation}
$ f_{2}^{S,0} = T_F C_{F} \left[ - \ln(\frac{M_H^2}{m_t^2}) + \frac{14}{3}
               + \left\{ -\frac{1}{6}\pi^2 -\frac{2}{3}
                           +\frac{1}{6}\ln^2(\frac{m_b^2}{M_H^2} )
                        \right\} \right]$,
\\ \\
$ f_{2}^{S,1} = T_F C_{F} \left[ - \frac{41}{1080} \ln(\frac{M_H^2}{m_t^2})
                +\frac{2011}{16200} +\left\{
                     - \frac{7}{720}\pi^2 - \frac{7}{180}
                              + \frac{7}{720}
			      \ln^2(\frac{m_b^2}{M_H^2} ) \right\}
    \right]$, \\ \\
$ f_{2}^{S,2} = T_F
C_{F} \left[ - \frac{47}{15120} \ln(\frac{M_H^2}{m_t^2}) +
			   \frac{28307}{3175200} + \left\{
			-\frac{1}{1008} \pi^2 - \frac{1}{252}
                           +\frac{1}{1008} \ln^2(\frac{m_b^2}{M_H^2} )
      \right\} \right]$, \\ \\
$ f_{2}^{S,3} = T_F
C_{F} \left[ - \frac{59}{168000} \ln(\frac{M_H^2}{m_t^2}) +
                        \frac{100381}{105840000} + \left\{
                          - \frac{13}{100800}\pi^2 - \frac{13}{25200}
                           + \frac{13}{100800} \ln^2(\frac{m_b^2}{M_H^2} )
                       \right\} \right]$,
\\ \\
where we presented separately the contributions from the non-singlet
diagrams, $\Gamma^{NS}(H\rightarrow b\overline{b})$,
and from the singlet diagram, $\Gamma^{S}(H\rightarrow b\overline{b})$.
The contribution of the two gluon cut is subtracted from the singlet
part (the corresponding terms are indicated in curly
brackets). $G_F$ is the Fermi constant.
$C_{F} = \frac{4}{3}$ and $C_{A}= 3$
are the Casimir operators of the fundamental and adjoint
representation of the colour group $SU(n)$, n = 3 is the number of
quark colours, $T_{F} = \frac{1}{2}$ is the trace normalization of
the fundamental representation.  The result is for 6-flavour QCD,
$N_f = 6$, $\alpha_{s}^{(6)}$ is the strong coupling constant and
$m_b^{(6)}(\mu)$, $m_t(\mu)$ are running $\overline{MS}$ bottom and
top quark masses. The effects of using the pole $m_b$ mass instead
of the running $m_b$ mass can be found in \cite{kataev}.

The singlet type coefficients $f_2^{S,0}$ and $f_2^{S,1}$ agree
with the previous calculations \cite{bbbarsin1}
and \cite{bbbarsin2} correspondingly. The
non-singlet type coefficients
agree with the previous calculation \cite{bbbarns}. The singlet
coefficients $f_2^{S,2}$, $f_2^{S,3}$ are new results.

To obtain the result in effective 5-flavour QCD where the non-singlet
top quark contributions decouple \cite{appelquist}, one should substitute
the b-quark mass $m_b^{(6)}$ and the coupling constant $\alpha_{s}^{(6)}$ in
terms of their values in the effective 5-flavour QCD by appling the known
decoupling relations \cite{ovrut,bernreuther,zbbig} :
\begin{equation} \label{decouplingrelation}
 \frac{\alpha_{s}^{(6)}(\mu)}{\pi} = \frac{\alpha_{s}^{(5)}(\mu)}{\pi}
 +\left( \frac{\alpha_{s}^{(5)}(\mu)}{\pi} \right)^{2} \frac{T_{F}}{3}
\ln(\frac{\mu^2}{m_{t}^{2}(\mu)})  + O(\alpha_{s}^{3}) \hspace{4.2cm}
\end{equation}
\[
 m_{b}^{(6)}(\mu) = m_{b}^{(5)}(\mu) \left[
   1 + \left( \frac{\alpha_{s}^{(5)}(\mu)}{\pi} \right)^{2} T_F C_F
        \left( - \frac{1}{8}\ln^2(\frac{\mu^2}{m_{t}^{2}(\mu)})
 \right. \right. \hspace{4cm} \]
\begin{equation} \label{decouplingmass}
\left. \left. \hspace{4cm}
     + \frac{ 5 }{24} \ln(\frac{\mu^2}{m_{t}^{2}(\mu)})
                 -\frac{89}{288}  \right) + O(\alpha_{s}^{3}) \right]
\end{equation}
Substitution of (\ref{decouplingrelation}) and (\ref{decouplingmass})
in (\ref{bbnsfull})+(\ref{bbsfull}) and
putting $\mu = M_H$ gives the following result
for the Higgs decay rate into $b\overline{b}$ in effective
5-flavour QCD
\begin{equation}
\Gamma(H\rightarrow b\overline{b}) =
\frac{3 G_{F}M_{H}(m_b^{5})^2}{4\pi\sqrt{2}}
\left[ 1+ \left( \frac{\alpha_{s}^{(5)}}{\pi} \right)  f_{1} +
\left( \frac{\alpha_{s}^{(5)}}{\pi} \right)^{2}
 f_{2} + O\left( \frac{\alpha_{s}^{(5)}}{\pi} \right)^{3}
     \right] , \hspace{4cm}
\end{equation}
$f_{1} = \frac{17}{3} \approx 5.66667 $,
\\ \\
$f_2 =  \frac{9235}{144} - \frac{97}{6}\zeta_3
          - \frac{17}{12}\pi^2 - \frac{2}{3} \ln(x)  + \frac{1}{9} \ln^2(y) $

$ + \left[ \frac{5233}{24300} - \frac{7}{1080} \pi^2
              - \frac{113}{1620} \ln(x) + \frac{7}{1080} \ln^2(y) \right] x  $

$ + \left[ - \frac{1837}{680400}  - \frac{1}{1512} \pi^2
          + \frac{1}{3240}  \ln(x) + \frac{1}{1512} \ln^2(y) \right] x^2  $

$ + \left[ \frac{3411287}{4286520000} - \frac{13}{151200} \pi^2
          - \frac{3193}{6804000}\ln(x) + \frac{13}{151200} \ln^2(y) \right] x^3
  + O(x^4)  \vspace{.1cm} $ \\

$  \approx 30.71675  - 0.66667 \ln(x) + 0.11111\ln^2(y) $

$ + \left[ 0.15138 - 0.06975 \ln(x)  + 0.0064815 \ln^2(y) \right] x $

$ + \left[ - 0.009227 + 0.00030864 \ln(x) + 0.00066137 \ln^2(y) \right] x^2 $

$ + \left[  -0.00005276  - 0.0004693 \ln(x) + 0.00008598 \ln^2(y) \right] x^3
           + O(x^4) $
\\ \\
where, $x = M_H^2/m_t^2$, $y = m_b^2/M_H^2$.

Please note
in the leading order of the large top quark mass expansion
the presence of $\ln(M_H^2/m_t^2)$ coming from the singlet
contribution which is not affected by the decoupling procedure
(since the singlet type contributions are renormalized independently
of the non-singlet ones).
It shows the violation of decoupling
for the considered process due to the singlet contribution,
which was first found in \cite{bbbarsin1,bbbarsin2}.

We conclude this section with stressing the fast
convergence of the obtained large top quark mass expansion for
$\Gamma(H\rightarrow b\overline{b})$
in the region below the threshold of $t \overline{t} $ production, $M_H<2m_t$
as can be seen from the fast decrease of the coefficients of the expansion.
Even in the region close to this threshold (where $x\approx 4$) one observes
%(expressing the expansion is terms of a more appropriate variable
% $M_H^2/(4 m_t^2)$)
a fast convergence of the large top quark mass expansion.

\section{The order $\alpha_{s}^3$ corrections to
                          $\Gamma(H\rightarrow$ gluons) }

Higgs boson decay into gluons is possible
since both the Higgs particle and gluons couple to massive quark loops.
An important property of this partial decay rate is that contributions
from heavy quarks in these loops are not suppressed by their mass
such that measurement of this decay rate counts the number of heavy quarks.
The contributions to $\Gamma(H\rightarrow$ gluons)
start from the order $\alpha_{s}^2$ (and this process is therefore
not as prominent phenomenologically
as the Higgs decay rate into bottom quarks which receives
tree level contributions).

$\Gamma(H\rightarrow$ gluons) is known\cite{japan,hgluons} in the
order $\alpha_{s}^3$ in the leading order of the large top quark mass
expansion and the $\alpha_{s}^3$ correction
turns out to be large (about 2/3 of the leading
$\alpha_{s}^2$ contribution).

In this section we present the higher orders in the large top quark
mass expansion of the Higgs decay rate into (two and three) gluons
$H \rightarrow gg(g)$
mediated by top quark loops.
The $gq\overline{q}$ final states
(with light quarks q=u,d,s,c,b) generated in the
chain $H\rightarrow gg \rightarrow gq\overline{q}$ are also included.
For this process we work in the approximation in which the
first five quark flavours are massless.

By applying the optical theorem,
the necessary contributions are given by the imaginary part of
the singlet 4-loop diagrams of the Higgs propagator
in which each of the external
Higgs vertices is located in a different top quark loop.
Clearly, for a Higgs mass below the top threshold, $M_H < 2 m_t$,
the only physical cuts that can be drawn in these propagator type diagrams
are through gluons and light quarks.

All contributing diagrams are of the singlet type and were
encountered previously
in the calculation of $\Gamma(Z\rightarrow$ hadrons) \cite{zbbig}
where a complete list of the necessary 4-loop singlet
diagrams was given.
Applying the diagrammatic large quark mass expansion to all contributing
diagrams yields the result in six flavour QCD, $N_f=6$,

 \[
\Gamma(H\rightarrow \mbox{gluons}) =
\frac{G_{F}M_{H}^3}{72\pi\sqrt{2}}
\left( \frac{\alpha_{s}^{(6)}}{\pi} \right)^{2}
  T_F^2 D
\left\{
 1 + h_{2}^{1}\frac{M_H^2}{m_{t}^2}
          +h_{2}^{2}\frac{M_H^4}{m_{t}^4}
 + O\left( \frac{M_H^6}{m_{t}^6}\right)
\right.
\hspace{6cm} \]
\begin{equation} \label{higgsgluons}
\left.  + \left( \frac{\alpha_{s}^{(6)}}{\pi} \right)
\left[ h_{3}^{0} + h_{3}^{1}\frac{M_H^2}{m_{t}^2}
                 + h_{3}^{2}\frac{M_H^4}{m_{t}^4}
                    + O\left( \frac{M_H^6}{m_{t}^6}\right)  \right]
\hspace{.05cm} \right\} ,
\end{equation}
$ h_{2}^1 = \left( \frac{7}{60} \right) $,
\\ \\
$ h_{2}^2 =  \left( \frac{1543}{100800} \right) $,
\\ \\
$ h_{3}^0 =  C_A \left[ \frac{103}{12}
                   - \frac{11}{6} \ln(\frac{M_H^2}{\mu^2}) \right]
            +  C_F \left( -\frac{3}{2} \right) $

        $   + T_F N_f \left[ - \frac{7}{3}
                      + \frac{2}{3} \ln(\frac{M_H^2}{\mu^2}) \right]
            + T_F \left[ \frac{7}{3}
                      - \frac{2}{3} \ln(\frac{M_H^2}{m_t^2}) \right] $,
\\ \\
$ h_{3}^1 = C_A \left[ \frac{71}{80}
                     - \frac{77}{360} \ln(\frac{M_H^2}{\mu^2}) \right]
          +  C_F \left[ \frac{13}{720}
                     - \frac{7}{40} \ln(\frac{\mu^2}{m_t^2}) \right] $

 $         + T_F N_f \left[ -\frac{29}{120}
                           + \frac{7}{90} \ln(\frac{M_H^2}{\mu^2}) \right]
           + T_F \left[ \frac{29}{120}
                         - \frac{7}{90} \ln(\frac{M_H^2}{m_t^2}) \right] $,
\\ \\
$ h_{3}^2 = C_A \left[ \frac{47459}{432000}
                      - \frac{16973}{604800}\ln(\frac{M_H^2}{\mu^2}) \right]
          + C_F \left[ \frac{83}{10080}
                    - \frac{1543}{33600} \ln(\frac{\mu^2}{m_t^2}) \right] $

 $         + T_F N_f \left[ -\frac{89533}{3024000}
                       + \frac{1543}{151200} \ln(\frac{M_H^2}{\mu^2}) \right]
           + T_F \left[ \frac{89533}{3024000}
                      - \frac{1543}{151200} \ln(\frac{M_H^2}{m_t^2}) \right] $,
\\ \\
where $D = n^2-1$ is the number of generators
of the colour group $SU(n)$ ($D=8$ for QCD).
In the leading order of the large top quark mass expansion
our $\alpha_s^3$ result agrees
with \cite{hgluons}.  Explicit checks show that the coefficients of
the logarithms in eq. (\ref{higgsgluons}) are in agreement with the
required renormalization group invariance of the physical quantity
$\Gamma(H\rightarrow \mbox{gluons})$.
Furthermore, the results that are presented
in this paper were obtained in an arbitrary
covariant gauge for the gluon fields i.e. keeping the gauge parameter as a
free parameter in the calculations.
The explicit cancellation of the gauge dependence
in the physical quantities gives a good check of the results.

Applying the decoupling relation (\ref{decouplingrelation}),
putting $\mu = M_H$ and substituting the QCD colour factors
gives the following result for effective 5 flavour QCD

\begin{equation}
\Gamma(H\rightarrow \mbox{gluons}) =
\frac{G_{F}M_{H}^3}{36 \pi\sqrt{2}}
 \left( \frac{\alpha_{s}^{(5)}}{\pi} \right)^{2}
\left[
 h_{2}  +\left( \frac{\alpha_{s}^{(5)}}{\pi} \right) h_{3}
 + O( \alpha_{s})^2 \right] ,
 \hspace{7cm} \end{equation}
$h_2 = 1 + \frac{7}{60} x + \frac{1543}{100800} x^2 + O(x^3) \vspace{.2cm}$

$\approx 1 + 0.116667 x + 0.015308 x^2 + O(x^3) $,
\\ \\
$h_3 =\frac{215}{12} + \left[ \frac{2249}{1080} - \frac{7}{30}\ln(x) \right] x
  + \left[ \frac{1612013}{6048000} - \frac{1543}{25200} \ln(x) \right] x^2
+ O(x^3)  \vspace{.2cm} $

$ \approx 17.9167
      + \left[ 2.08241 - 0.23333 \ln(x) \right] x
           + \left[ 0.26654 - 0.06123 \ln(x) \right] x^2 + O(x^3) $.
\\ \\
where $x= M_H^2/m_t^2$.

For completeness we will also give the total hadronic decay rate
of the Higgs boson (in the leading order of the small b-quark mass
expansion).  This result is obtained by summing the partial decay
rates $\Gamma(H\rightarrow b\bar{b})$ and $\Gamma(H\rightarrow
$gluons) i.e. by applying the decoupling relations to the sum of eqs.
(\ref{bbnsfull}), (\ref{bbsfull}) and (\ref{higgsgluons}) without
subtracting the two gluon cut in eq.(\ref{bbsfull}).
%\pagebreak
\[
\Gamma_{\rm tot}(H\rightarrow \mbox{hadrons}) = \hspace{12cm}
\vspace{.1cm} \] \[ \frac{3 G_{F}M_{H}(m_b^{(5)})^2}{4\pi\sqrt{2}}
\left[ 1+ \left( \frac{\alpha_{s}^{(5)}}{\pi} \right)  \frac{17}{3} +
\left( \frac{\alpha_{s}^{(5)}}{\pi} \right)^{2} \left\{
\frac{9299}{144} - \frac{97}{6}\zeta_3 - \frac{47}{36}\pi^2 -
 \frac{2}{3} \ln(x) \right.  \right.  \hspace{3cm}
 \]
\[
%\hspace{1cm}
 + \left[ \frac{5863}{24300} -
 \frac{113}{1620} \ln(x)  \right] x +\left.  \left[ -
 \frac{37}{680400} + \frac{1}{3240}  \ln(x)  \right] x^2 + O(x^3)
 \right\} +O(\alpha_s^3) +O(m_b^2)  \Biggr] \] \[ +
\frac{G_{F}M_{H}^3}{36 \pi\sqrt{2}} \left[ \left(
 \frac{\alpha_{s}^{(5)}}{\pi} \right)^{2} \left[ 1 + \frac{7}{60} x +
\frac{1543}{100800} x^2 + O(x^3)  \right] + \left(
 \frac{\alpha_{s}^{(5)}}{\pi} \right)^{3} \left\{ \frac{215}{12}
 \right. \right. \hspace{6cm} \] \[
%\hspace{2.4cm}
 \left. + \left[
 \frac{2249}{1080} - \frac{7}{30}\ln(x) \right] x + \left[
  \frac{1612013}{6048000} - \frac{1543}{25200} \ln(x) \right] x^2 +
O(x^3) \right\} +O(\alpha_s^4) +O(m_b^2) \Biggr] \hspace{.4cm}
\vspace{.1cm}
\] \[ \]
\[  \approx \frac{3 G_{F}M_{H}(m_b^{(5)})^2}{4\pi\sqrt{2}}
\left[ 1+ \left( \frac{\alpha_{s}^{(5)}}{\pi} \right)  5.66667 +
\left( \frac{\alpha_{s}^{(5)}}{\pi} \right)^{2} \left\{ 32.2578
 -0.6667 \ln(x) + \biggl[ 0.2413  \right.  \right.  \hspace{3cm}
 \]
\[
%\hspace{1cm}
    -
 0.06975 \ln(x)  \biggr] x +\left.  \biggl[ -
 0.00005438 + 0.0003086 \ln(x)  \biggr] x^2 + O(x^3)
 \right\} +O(\alpha_s^3) +O(m_b^2)  \Biggr] \] \[ +
\frac{G_{F}M_{H}^3}{36 \pi\sqrt{2}} \left[ \left(
 \frac{\alpha_{s}^{(5)}}{\pi} \right)^{2} \biggl[ 1 + 0.11667 x +
 0.01531 x^2 + O(x^3)  \biggr] + \left(
 \frac{\alpha_{s}^{(5)}}{\pi} \right)^{3} \biggl\{ 17.9167
 \right.  \hspace{6cm} \] \[
%\hspace{2.4cm}
 \left. + \biggl[
 2.0824  - 0.2333\ln(x) \biggr]  x + \biggl[
  0.2665  - 0.06123 \ln(x) \biggr] x^2 +
O(x^3) \right\} +O(\alpha_s^4) +O(m_b^2) \Biggr] , \hspace{2cm}
\vspace{.1cm}
\]

where $x= M_H^2/m_t^2$. Here the terms proportional to $m_b^2$
originate from $\Gamma(H\rightarrow b \overline{b})$
and the terms proportional
to $M_H^3$ come from $\Gamma(H\rightarrow$ gluons)
(both in the leading order of the small $m_b$-expansion).

We
conclude that the large top quark mass expansion converges rapidly
for $\Gamma(H\rightarrow b \overline{b})$,
$\Gamma(H\rightarrow$ gluons) and $\Gamma_{\rm tot}(H\rightarrow
\mbox{hadrons})$ in the region of its applicability
$M_H<2m_t$.

\section{Acknowledgements}
We are grateful to K.G. Chetyrkin, A.L. Kataev,
R. Kleiss, J. Smith and M. Veltman for helpful discussions.
S.A. Larin gratefully acknowledges the support of the NWO
project "Computer Algebra and Subatomic Physics"
and of the Russian Fund for Fundamental Research, Grant No.
94-02-04548-a.

\end{document}